# Assessing the differences between numerical methods and real experiments for the evaluation of reach envelopes of the human body

Mathieu Delangle
IRCCyN, *Ecole Centrale de Nantes*
*1, rue de la Noe, 44321, Nantes, France*
Mathieu.delangle@irccyn.ec-nantes.fr

Jean-François Petiot and Emilie Poirson
IRCCyN, *Ecole Centrale de Nantes*
*1, rue de la Noe, 44321, Nantes, France*
Jean-François.Petiot@irccyn.ec-nantes.fr
emilie.poirson@irccyn.ec-nantes.fr

*Résumé*— The use of static human body dimensions to assess the human accessibility is an essential part of an ergonomic approach in user-centered design. Assessments of reach capability are commonly performed by exercising external anthropometry of human body parts, which may be found in anthropometric databases, to numerically define the reach area of an intended user population. The result is a reach envelope determined entirely by the segment lengths, without taking into account external variables, as the nature of the task or the physical capacities of the subject, which may influence the results. Considering the body as a simple assembly of static parts of different anthropometry is limiting. In this paper, the limit of validity of this approach is assessed by comparing the reach envelopes obtained by this method to those obtained with a simple two-dimensional experimental reaching task of a panel of subjects. Forty subjects experimentally evaluated the reach, first with the body constrained and second unconstrained. Results were recorded and compared with those obtained numerically with a model, based on their own anthropometric characteristics, previously measured. A statistical study of the results allowed the definition of the shape of a confidence bound containing the real reach envelope. The results indicated important differences between the experiment and the numerical evaluation of the reach envelope.

I. INTRODUCTION

Modeling normal human reach is widely used by the designer to design and assess the accessibility of environments. Many tools and practices are based on anthropometrics data to perform these ergonomics evaluations [1]. Appropriate anthropometric data regarding body size from the established data bases are used to analyze and design the intended product or environment. Existing database (e.g. ANSUR [2], NHANES [3]) are generally chosen as the reference population and are used directly to assess the environment, instead of recruiting sample-users to test the product. However, the reach behavior on an individual depends on many factors. Anthropometry, age, gender, joint mobility and muscle strength are a few such factors related to the individual being modeled. This human variability might cause difficulty to meet the requirements of a conception, especially for a certain percentage of the user population.

Although numerical methods are faster and less expensive than the involvement of sample-user to test prototypes, the only use of these anthropometric data [4], often old, are not always representative of the target users population. In fact, they often consist of specifics surveys (military...), and typically provide only very limited information concerning children and people who are older and disabled [5]. Moreover, these data are generally used in univariate case study (to determine the appropriate allocation of adjustability to achieve a desired accommodation level), where most problems are multidimensional. That is why, although design methods based on external body dimensions don't need experimental tests or the building of prototypes, methods based on this principle still pose questions about their ease of use and their reliability compared to reality.

This paper describes to which extent an evaluation of the reach only based on the structural data of the human body may differ from those obtained with an experimental task. Thus, an experimental reach assessment and a numerical evaluation were performed and compared to highlight differences.

II. COMPARATIVE STUDY

The present study proposes to compare two ways of accessibility assessment of the human body. First during an experiment, a sample-user is asked to perform an accessibility task. Second using numerical data and a kinematic model of the body, the structural data of the participant are directly used to numerically assess the accessibility (Figure 1).

The task proposed is a 2 dimensional reaching task of the hand of the participant in the frontal plane (Figure 2). Experimental test was divided into two sub-tests; a constrained test (feet fixed to the floor and body fixed relative to the vertical axis of the center of the plate) and an unconstrained test (feet fixed to the floor and rest of body free to move). The unconstrained situation represents the functional reach, that is to say the maximal distance one can reach with the hand while maintaining a fixed base of support in the standing position [6], [7].

## III. EXPERIMENT

### A. Sampling

The experiments were conducted with 40 adult volunteers, all French students or teachers of the Ecole Centrale de Nantes.

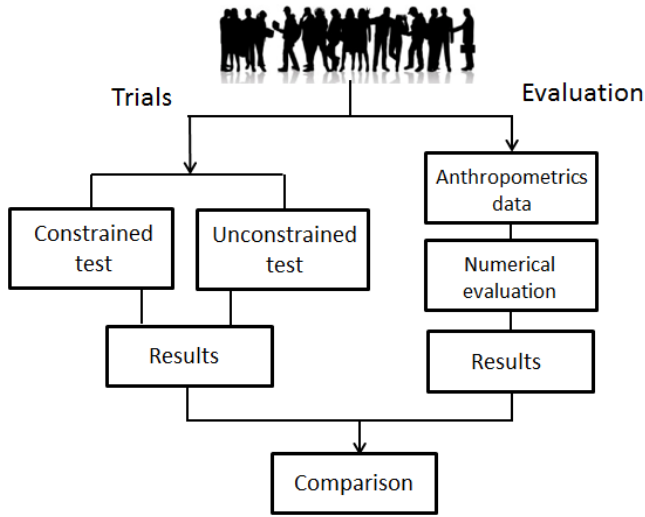

Figure 1.    SYNOPSIS OF THE STUDY

Twenty five males and fifteen females were sampled in the study, covering a wide spectrum of physical characteristics, from 1482 mm for the smallest stature, to 1930 mm for the highest. The average of stature of the subjects was 1735 mm (with a standard deviation of 95.4). The summary statistics of the anthropometry of the subjects are provided in Table 1. No one of these volunteers reported motor disabilities or particular physiological limitations. The sample is considered as representative of the general population. It was a deliberate decision to not skew the data by "excluding" persons in the panel (e.g. old or disabled), in order to not bias the comparison. Indeed, the accessibility study of persons with specific physical limitations will depend of more parameters that would make the comparison more difficult.

|  | 1 | 2 | … | 40 | Mean | S.D. |
|---|---|---|---|---|---|---|
| Gender | M | F | … | M | - | - |
| Shoulder height | 1485 | 1450 | … | 1420 | 1455 | 87.7 |
| Shoulder width | 470 | 430 | … | 460 | 456 | 39.8 |
| Arm length | 750 | 730 | … | 710 | 719 | 41.5 |
| Stature | 1735 | 1705 | … | 1715 | 1735 | 95.4 |

Table 1.    ANTHROPOMETRIC CHARACTERISTICS OF THE SUBJECTS (IN MM) PARTICIPATING TO THE EXPERIMENT, WITH THE AVERAGE AND THE STANDARD DEVIATION (EXCERPT).

### B. The tests

*Principle*: Marking of the boundaries of the hand reach in standing posture.

The boundary of reach volumes based on the reach capabilities of both arms were measured based on a technique used in [8], [9] and [10]. Briefly, a white board was fixed vertically, and individuals were asked to produce an 'arc' averaging 90° (from the shoulder height to the top of the head). In order to check if the subjects does not lift their feet to increase their vertical reach, an electronic sensor position was positioned under the heels indicating if the feet are off the ground or not. When the visual signal was triggered, the test was stopped and the subject repositioned. For each individual, the reached envelope was drawn on a white board. Arcs were drawn with the body, providing the volume described by the reach capability for both arms. Envelopes were marked to represent the task as "finger touch" function (one finger touches an object without holding it) in order to avoid much as possible grasp effects (reducing the reach envelope). Because most anthropometric data presented in databases represent nude body measurements and to permit reliable comparison, experiments were performed with light clothing (nude dimension and light clothing being regarded as synonymous for practical purposes).

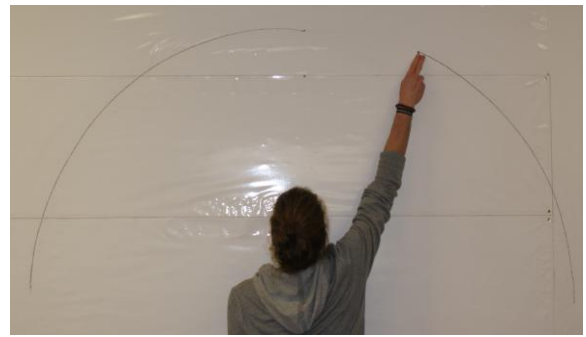

Figure 2.    PARTICIPANT DRAWING HIS REACH ENVELOPE DURING THE CONSTRAINED TEST.

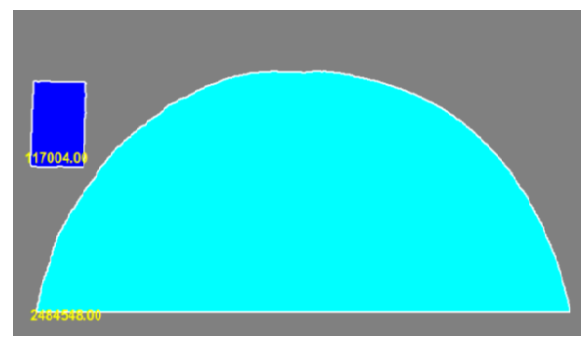

Figure 3.    IMAGE PROCESSING USING MATLAB®. BRIGHT SPACE REPRESENTS THE REACH ENVELOPE AREA DRAWN BY THE PARTICIPANT. THE DARK AREA (RECTANGLE) REPRESENTS THE PAPER SHEET USED AS REPOSITORY TO CALCULATE AREAS FROM PIXELS TO SQUARE METERS.

Figure 2 shows a subject performing the reach envelope task using the traditional reach envelope board method. The arcs produced were captured on photo for analysis; pictures were treated with a MATLAB® program to accurately determine the area of the envelope (Figure 3). Each subject performed this test firstly constrained (foot and body fixed) and secondly unconstrained (only foot fixed). A similar protocol where participants are in interaction with physical points may be found in [**11**].

## IV. STATIC NUMERICAL EVALUATION

### A. Creation of the anthropometric database

When using methods for ergonomic evaluations, anthropometric modeling can be either directly observed from anthropometric characteristics of the current users, or statistically derived from characteristics for the intended target population. In order to predict the accessibility by limiting the statistical biases in the comparison, accurate anthropometric data from the participants were needed. Thus, the presented evaluations were all based on the anthropometric characteristics of the subjects who performed the experimental tests. External anthropometry being the type most frequently available and collected, it was decided to collect some "direct" measurements of external link-length anthropometry. Data in Table 1 were recorded to predict the upper body accessibility for each participant, and were collected in a laboratory environment from the 40 individuals. It was expected that this number would provide a manageable database for the development and validation of the comparative study.

### B. The numerical model

The aim is to evaluate the reach characteristics from the recorded external anthropometrics that might be found in anthropometric database (Table 1). This methodology is based on the design limits approach, which is a common method used in design problems, where data about human physical characteristics are directly applied to solve design problem. The maximal reach is directly correlated to the greatest distance between the shoulder acromion and the fingertips, corresponding to an outstretched arm situation. This is kinematically modeled as a simple link (arm length) with a unique revolute joint (shoulder). So, the maximal reach envelope is defined by an arc circle, with a radius equal to the arm length of the operator and the shoulder as point of rotation (Figure 4). All points within this envelope (shaded area) were considered as reachable by the subject. The total reach area was defined using Equation 1. H represents the tip of the hand, S the shoulder location and O the center of the shoulder width, thus OS the half shoulder width and SH the arm length. C1 represents the arc circle area between H and S, C2 the arc circle area between S and O, and C3 the area of the SOP triangle.

Knowing the anthropometric characteristics of each participant, a program was implemented (using MATLAB® R2012b) allowing to automatically determinate the corresponding reach. The results are presented in section V.

Eq.(1) :
$$A_{total} = 2.(C_1 + C_2 + C_3) = 2.\int_S^H \sqrt{SH^2 - x^2}.dx + 2.\int_S^O \sqrt{SH^2 - x^2}.dx + OS.OP$$

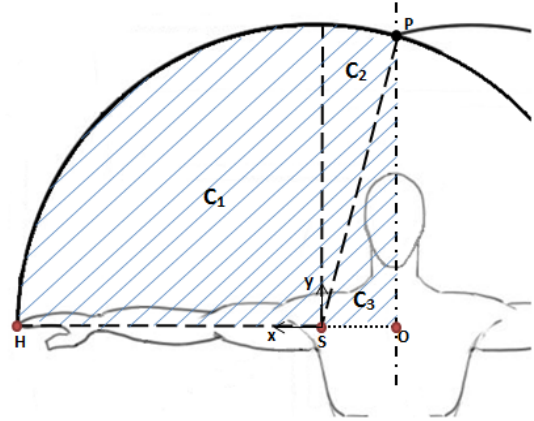

Figure 4. STATIC MODEL OF REACH ENVELOPE (SHADED AREA) OF THE RIGHT ARM CALCULATED FROM EXTERNAL ANTHROPOMETRICS DIMENSIONS OF THE SUBJECT.

## V. RESULTS

Results from experiments and evaluations were collected for each of the forty participants. The aim is to compare the shape of the reach envelopes obtained from the experimentation with those obtained from numerical assessments. Obviously, the anthropometric characteristics of the participants being different, the shapes of the envelopes are all different too. So as to make an objective comparison of the general behavior for the forty profiles, a normalization procedure was performed to compare the reaches between all the participants. First, experimental reach envelopes were recorded and drawn for each subject. Second, each envelope was rescaled in a common domain of comparison (1 unit representing the arm length of each participant). Thus, for all of them, the envelope defined by the numerical model (section IV.B) is modeled by a common arc circle, where 0 corresponds to the shoulder position and 1 to the theoretical extremity of the hand for a static position. This normalized theoretical reached envelope was used as a basis for comparisons to highlight the differences between the real reaches and the numerical model. Finally, the forty normalized reaches obtained from the experiments were aggregated on the same graph and the extreme boundaries (corresponding to the extreme position of fingertips) were extracted. The average extreme reach obtained for the sample-users is depicted Figure 5.

The sample of population being considered as representative of a normal distribution, statistical evaluations were made to represent, depending on the position of the arm, the variability of reach. This variability is the differences between the theoretical curve based on the angle of the arm, and the calculation of an overall confidence reach interval (95% of the population). Given the sample size "n" (>30), it is estimated that the mean x of the sample follows a normal distribution. Thus, the confidence interval of 95% could be determinate using Equation 2. Where n is the size of the sample, σ the standard deviation and x the mean. The results of the differences between the experimental continuous reaches and those numerically found are presented Table 2 for the constrained test and Table 3 for the unconstrained. The associated graphic representation is shown Figure 5.

Eq. (2):  $$I = \overline{x} \pm \frac{1.96*\sigma(s)}{\sqrt{n}}$$

|  | 0° | 10° | 30° | 45° | 60° | 80° | 90° |
|---|---|---|---|---|---|---|---|
| $\overline{x}$ | 1.086 | 1.059 | 1.038 | 1.016 | 1.008 | 0.995 | 1.003 |
| $\sigma$ | 0.073 | 0.07 | 0.062 | 0.056 | 0.055 | 0.06 | 0.058 |
| $I^+$ | 1.11 | 1.08 | 1.06 | 1.03 | 1.03 | 1.02 | 1.02 |
| $I^-$ | 1.062 | 1.04 | 1.02 | 1.00 | 0.99 | 0.98 | 0.98 |
| $I^+-I^-$ | 0,047 | 0,046 | 0,041 | 0,037 | 0,036 | 0,039 | 0,038 |

Table 2.   RESULTS OF THE CONSTRAINED TASK: POSITIONS OF THE HAND FOR SEVERAL ARM ANGLES, AFTER AGGREGATION OF THE 40 CONTINUOUS REACH ENVELOPES; WITH X THE MEAN, Σ THE STANDARD DEVIATION, $I^+$ AND $I^-$ RESPECTIVELY THE SUPERIOR AND INFERIOR LIMITS OF THE 95% CONFIDENCE INTERVAL.

|  | 0° | 10° | 30° | 45° | 60° | 80° | 90° |
|---|---|---|---|---|---|---|---|
| $\overline{x}$ | 1.363 | 1.317 | 1.241 | 1.181 | 1.127 | 1.081 | 1.076 |
| $\sigma$ | 0.104 | 0.098 | 0.077 | 0.073 | 0.065 | 0.056 | 0.055 |
| $I^+$ | 1.397 | 1.349 | 1.266 | 1.205 | 1.148 | 1.1 | 1.094 |
| $I^-$ | 1.329 | 1.285 | 1.215 | 1.157 | 1.106 | 1.063 | 1.058 |
| $I^+-I^-$ | 0,068 | 0,064 | 0,050 | 0,048 | 0,042 | 0,037 | 0,036 |

Table 3.   RESULTS OF THE UNCONSTRAINED TASK: RESULTS OF POSITIONS OF THE HAND FOR SEVERAL ARM ANGLES, AFTER AGGREGATION OF THE 40 CONTINUOUS REACH ENVELOPES; WITH X THE MEAN, Σ THE STANDARD DEVIATION, $I^+$ AND $I^-$ RESPECTIVELY THE SUPERIOR AND INFERIOR LIMITS OF THE 95% CONFIDENCE INTERVAL.

The 95% interval indicates that there is 95% chance that the interval contains a new observation. It can be seen that, for the constrained and unconstrained experiments, the means curves, not perfectly follow the theoretical envelope numerically defined.

Looking to the shape of the constrained experiment, it can be seen that the vertical reach is perfectly coincident with the theoretical results. The mean (with the corresponding interval confidence) is 1±0.02. So, for this situation, the experiment is coherent with the numerical model based on the structural body dimensions. For the lateral reach, the results show that the experimental reach is greater than those of the theoretical model, with an average of 1.086±0.024.

Results obtained for the unconstrained show that the reach is overall more important, compared to the numerical model and to the constrained experiment. The normalized mean is 1.076±0.018 for the vertical location, and 1.363±0.034 for the lateral reach.

In the constrained case, the extended lateral reach is certainly due to small displacements during the test. Indeed, although this experiment was implemented in order to limit as much as possible movements of the body, displacements of the upper body could appear during the tests.

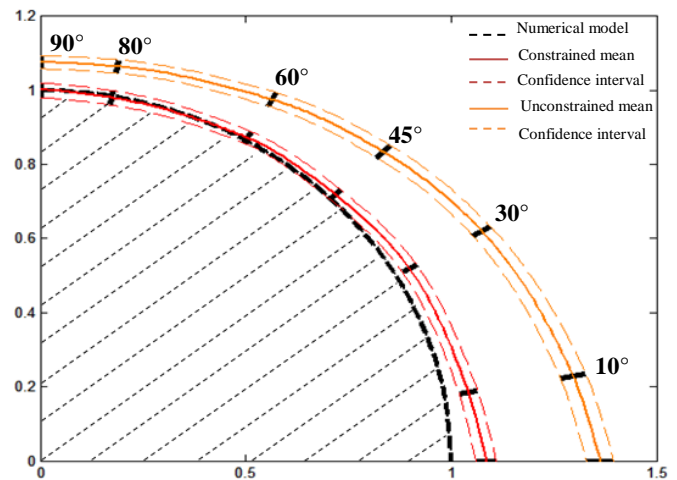

FIGURE 5.   NORMALIZED MEAN REACH LINES OBTAINED FROM THE EXPERIMENT, WITH THE 95 PERCENT CONFIDENCE INTERVAL. THE SHADED AREA REPRESENTS THE NORMALIZED THEORETICAL REACHED AREA OBTAINED FROM THE STATIC NUMERICAL EVALUATION.

For the unconstrained task, this extended lateral reach represents the realistic functional reach of the subject. The body being not constrained (excepted the feet), the maximal reach strongly depends on the lateral reach capacity, represented by the medio-lateral balance stability and the pelvis rotation ability of the subject, as depicted Figure 6. Thus, the maximal lateral reach is greatly higher than expected. Considering the average arm length, this average maximal lateral reach corresponds to a deviation of 26cm, which is consistent with results that can be found in the literature [12].

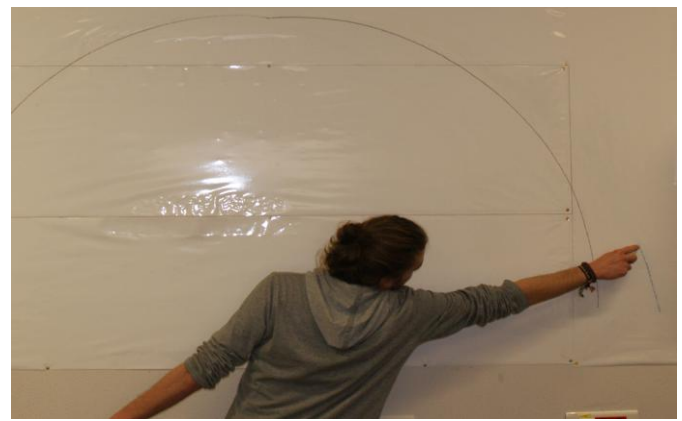

FIGURE 6.   PARTICIPANT EXTENDED HIS LATERAL REACH USING THE MEDIO-LATERAL BALANCE STABILITY AND THE PELVIS ROTATION, DURING THE UNCONSTRAINED TEST.

The vertical envelope obtained from the constrained experiment is perfectly coherent with the numerical model, which is not the case for the unconstrained test. This involve that the movement executed with a free body allows the participants to increase their vertical reach. The main reason of this increase is probably due to the complex model of displacement of the shoulder, and particularly of its flexion, allowing the subjects to increase their reach upwardly, beyond their external structural dimensions. With the foot fixed on the ground, the strategy to enhance the vertical reach is to increase the flexion of the shoulder and to change the alignment of the shoulder and the pelvis

The discrepancy between the constrained and the unconstrained reach is the highest for the lateral location, and decreases continuously up to the vertical. This represents the differences of strategy used throughout of the envelope to improve the maximum reach. The highest variability of the reach appears for the unconstrained test at 0°, with a standard deviation of σ=0.104. Indeed, the balance capacity of each subject might be very different, involving an important variation of the reach from one participant to another. This variations would be greater for a sample of people with very different physical characteristics (disabled persons, children…), which is not the case in this study. So, the biomechanics characteristics of the body directly impact the maximum reach capacity.

## VI. CONCLUSION

This study compared results of a reach assessment obtained from experiments and from numerical evaluations. In spite of the simple nature of the presented task (two-dimensional), the results show important differences between the two evaluations. The reach capacities of the participants were not only correlated to their anthropometric characteristics. The biomechanics of the body and the physical abilities of the human implies that the maximum reach change during the reach. The reach capacity does not represent a perfect arc circle but a curve depending of the hand position in the space.

Anthropometric design problems associated with human physical characteristics depend on many other factors, related to the task, the body position effects and the human behavior in interaction. Several task considerations should be taken into account in order to construct a reach envelope, as the nature and requirements of the task to be performed, the body position while reaching, the whole body movement capabilities and restraints. Moreover, numerous human variability factors might affect the reach results, as the age, gender, body build, fatigue, disease, clothing or environment. These parameters have to be taken into account in the reach evaluation, especially for multivariate design problem. Perform accessibility evaluations only considering the structural limits data of the problem can rapidly leads to misfit design solutions.

## VII. PERSPECTIVES

In continuation of this study, a methodology using three-dimensional human simulation is being investigated in order to incorporate real human reach capacity in virtual simulations, without the participation of sample-users. Based on CAD environment and virtual reality tools, reach assessment of multi-dimensional design problem will be performed in an intuitive and interactive manner. Thus, the external factors influencing the reach and the differences of physical capacities may be taken into account to carry out robust reach assessments, especially in the universal design field.